# Zirconia nano-colloids transfer from continuous hydrothermal synthesis to inkjet printing

M. Rosa [1*], P. N. Gooden [2], S. Butterworth [2], P. Zielke [1], R. Kiebach [1], Y. Xu [1], C. Gadea [1], V. Esposito [1]

[1]DTU Energy, Technical University of Denmark, Risø Campus, Frederiksborgvej 399, 4000, Roskilde, Denmark.

[2]Promethean Particles Ltd., Unit 1 Genesis Park, Midland Way, Nottingham, NG7 3EF, UK.

*corresponding author: masros@dtu.dk

**ABSTRACT**

Water dispersions of nanometric yttria stabilized zirconia (YSZ) particles synthesized by Continuous Hydrothermal Synthesis are transferred into nano-inks for thin film deposition. YSZ nanoparticles are synthesized in supercritical conditions resulting in highly dispersed crystals of 10 nm in size. The rheology of the colloid is tailored to achieve inkjet printability (Z) by using additives for regulating viscosity and surface tension. Inks with a wide range of properties are produced. A remarkable effect of nanoparticles on the ink printability is registered even at solid load < 1 %vol. In particular, nanoparticles hinder the droplet formation at low values of the printability while suitable jetting is observed at high Z values, *i.e.* Z ≈ 20. For the optimized inks, we achieve high quality printing with lateral and thickness resolutions of 70 μm and ca. 250 nm respectively, as well as self-levelling effect with a reduction of the substrate roughness. Densification is achieved at sintering temperatures below 1200 °C.

**Keywords**: Continuous Hydrothermal Synthesis, Nanomaterial, Zirconia, Inkjet, Thin Films

1. **Introduction**

Nanostructured materials show unique properties that can dramatically improve the performances of several technologies in many different fields, from catalysis [1] to optics [2], from optoelectronic [3] to structural applications [4]. From the application point of view, their higher activity compared to bulk materials is due to the combination of size effects [5] and high surface area [6,7]. The first modify the band structure of the material, affecting, *e.g.*, electronic and optical properties [5]. The latter has a direct impact on phenomena taking place at the interfaces, increasing the number of superficial defective sites [7]. As a consequence, nanoparticles typically show a decrease in the sintering temperature [8–13], which can be exploited in their processing. In addition, their nanometric size is ideal for deposition via liquid casting methods such as inkjet printing [14].

Inkjet printing is an additive manufacturing technique that consists in the accurate deposition of liquid droplets onto a substrate. By printing nanoparticles-loaded fluids, it is possible to fabricate customized ceramic patterns with high resolution, reproducibility and automation [15]. Typically, inkjet printing produces droplets with volumes in the picoliter range ($10^{-12}$ L), which are jetted from hundreds of nozzles with a diameter of a few tens of microns. Depositing picoliter-sized droplets allows controlling accurately the deposition of nanograms of particles, resulting in printing with high lateral resolution. Therefore, the formulation of finely dispersed nano-colloids for inkjet, i.e. nano-inks, paves the way for a further increase in the miniaturization capability of this technique. In particular, the reduced particle size allows decreasing the nozzle diameter and the droplet volume with a further enhancement in the lateral resolution. Moreover, thinner structures can be printed using smaller particles, increasing the thickness resolution.

However, using nanomaterials raises remarkable processing difficulties compared to micron-sized structures. One of the reasons for this is their large surface area, which usually leads to a strong increase of the inter-particle interaction, leading, eventually, to aggregation [16]. The preparation of monodispersed nano-colloids from dry powders is particularly challenging and often requires intense deagglomeration treatments and an extensive use of stabilizing agents.

The formulation of aqueous inks with suspended nanoparticles has already been successfully demonstrated for Ag colloids [17,18]. In particular, stable inks with a solid loading up to 45 % wt of Ag nanoparticles were prepared with the aid of a polyethilenoxide-polypropilenoxide co-polymer [17]. On the other hand, inkjet printing of several metal oxide nanoparticles ($Al_2O_3$, $TiO_2$, $WO_3$) was reported to be more challenging [19–23]. In particular, the ink formulation often required the use of low concentrated inks and a careful ink preparation for avoiding large agglomerates. For example, it was shown that dispersing commercial $TiO_2$ nanopowders involves an ultrasound treatment and a proper selection of the dispersant, despite a low particle content of 1-3 %wt [21,23]. Nonetheless, these efforts do not always lead to singularly dispersed nanoparticles, as a certain degree of agglomeration was observed in some cases [21,23,24]

One of the strategies to overcome these problems is synthesizing nanoparticles directly dispersed in a liquid medium. Continuous hydrothermal synthesis (CHS) represents a powerful route for the synthesis of various materials in the form of a liquid dispersion of nanosized particles with a narrow size distribution. An additional feature that increases the effectiveness of this method is operating in supercritical conditions [25–27]. In a typical CHS reactor, a stream of a precursor solution is continuously mixed with a stream of water above its critical point [28]. At the interface between the two flows, the precursor solution reaches rapidly supersaturation resulting in the abundant formation of nucleation sites. A large number of nanoparticles, for a capability in the range of tens of grams per minute, grow from these sites and are transported and collected downstream in an all-wet process. The reaction stoichiometry can be also controlled during synthesis by regulating the streams compositions and flows, providing constant synthesis conditions and uniform products. Moreover, CHS can be scaled-up [29] without handling dry nano-powders at any step of the processing.

Due to their highly dispersed particles, low viscosity and good scalability, CHS products have the potential to act as precursors for nano-inks for inkjet printing [14,30] On the other hand, controlling the rheology of a nano-dispersion is a non-trivial task, especially considering the extreme shear variations experienced by a fluid during inkjet printing [31,32]. The fluid dynamics of this process has been studied both from a theoretical [33–35] and experimental point of view [15,36,37]. Most of the experimental studies focus on particle-free fluids [36,38] or solid loaded inks with particles of hundreds of nanometers [39,40]. However, nano-fluids are known to have a different rheology than conventional fluids [41], which might affect inkjet printing [42].

In this work, we investigated the conversion of yttria stabilized zirconia (YSZ) nano-dispersions produced by continuous hydrothermal synthesis into nano-inks for inkjet printing. YSZ is one of the most studied ceramic ionic conductors and its properties are exploited in several applications and with different microstructures; *e.g.* dense electrolytes or porous backbones for electrodes and catalysts [43–45]. While inkjet printing opens to the possibility of fabricating application-tailored structures and apply novel designs [46,47], the colloidal properties of nano-dispersions need to be tailored after synthesis by CHS. Transferring CHS products to inkjet printing represents a possible strategy for taking advantage of nano-sized YSZ, which allows a lower sintering temperature [48] and a higher printing resolution.

2. **Experimental**

Two CHS reactors were utilized for the synthesis of YSZ nano-colloids in water. The reactor built at DTU Energy and the procedure for synthesis of YSZ had been described elsewhere [25]. Briefly, a precursor stream of 0.184 mol L$^{-1}$ Zr(NO$_3$)$_4$ and 0.032 mol L$^{-1}$ Y(NO$_3$)$_3$ with a flow rate of 10 mL min$^{-1}$ was mixed with a stream of supercritical water (equivalent to 25 mL min$^{-1}$ at room temperature) at 397 °C and 270 bar. The reactant solution was heated to approx. 390 °C after mixing before cooling down to room temperature. After collection, the products were washed and concentrated by centrifuging the dispersion, and redispersing particles in DI water.

The large scale YSZ production at Promethean Particles Ltd. started from yttrium nitrate and zirconium acetate as precursors, which were mixed with water in a feedstock solution. The precursor stream was pumped at 20 mL min$^{-1}$ in the reactor and mixed with a water flow in supercritical conditions at 375 °C and 241 Bar. The mixing zone is engineered with a high-pressure tube-in-tube counter current mixing system. Samples were depressurized and cooled down before increasing the particle concentration by tangential flow filtration [49]. Different batches were produced with a final solid loading in the range 5.5 ±1 %wt using the two reactors. The solid loading was evaluated by evaporating the solvent from the final dispersion. The as-produced particles were characterized by TEM, using a Jeol JEM 3000F microscope. The particle size distribution (PSD) of a synthesized YSZ dispersion was measured by Dynamic Light Scattering (DLS) with a Malvern Zetasizer Nano on a concentrated sample with a particle concentration of 6.5 %wt. The ζ-potential was measured on a synthesized YSZ dispersion after concentration with a solid loading of 4.5 %wt. The ζ-potential was monitored for 30 minutes at room temperature using a Zeta Probe Analyzer from Colloidal Dynamics. The concentrated dispersions showed little sedimentation after 72 hours on the shelf, and the solid deposit could be dispersed by gently stirring even letting the dispersion settle for more than 4 weeks.

Nano-dispersions produced by CHS were the starting materials for the formulation of the inks. The printability parameter Z was used to ensure printability and evaluate the properties of the inks, as reported in other papers [50]. This factor was first proposed by Fromm [33] as a figure of merit for describing the jetting behavior of a fluid. Z is inverse of the Ohnesorge number, $Oh$, which is independent from the velocity of the fluid and contains several properties of the ink in the form: $Z = 1/Oh = (\rho \cdot \sigma \cdot a)^{1/2} / \eta$ (1), where $\rho$ is the density, $\sigma$ is the surface tension, $\eta$ is the viscosity, and $a$ is the characteristic length, which is typically considered as the nozzle diameter [15].

Polyvinylpyrrolidone (PVP) K15 with an average molecular weight of 10K g mol$^{-1}$ (Applichem) and PVP with average molecular weight of 360K g mol$^{-1}$ (Sigma-Aldrich) were both used to regulate the viscosity of the ink. 2,4,7,9-Tetramethyl-5-decyne-4,7-diol ethoxylate (TMDE, Sigma-Aldrich) was used as a surfactant for controlling the surface tension. The pH was adjusted using HCl 36 %wt (Alfa Aesar). All the chemicals were used as received.

In a typical ink preparation, the CHS dispersions were first sonicated using a sonic probe (Hielscher UP200ST) with a 3 minutes burst at 50% of amplitude. Then, PVP and HCl were dissolved under magnetic stirring and before printing TMDE was added.

Viscosity was measured with a rheometer (Anton Paar, MCR 302) in rotational mode and at a constant temperature of 21 °C. Rheological measurements were carried out using a plate-plate system with a diameter of 50 mm and a gap distance of 0.6 mm. Analyzes were performed at increasing shear rates up to 1000 s$^{-1}$. The $\eta$ values at the highest shear rate were used for the calculation of the printability according to equation (1). Surface tension was measured with a bubble pressure tensiometer (BP50, KRÜSS).

Inks were jetted and printed using a Pixdro LP50 inkjet printer equipped with DMC disposable piezoelectric printheads from Dimatix. These printheads have 16 nozzles with a 21.5 μm diameter and a nominal droplet volume of 10 pL. For each ink, the waveform actuating the piezoelectric elements responsible for the droplet ejection was optimized starting from a standard trapezoidal pulse. This waveform was characterized by a dwell time and fall time of 10 and 5 μs respectively.

Optimized printing was achieved by changing the filling time for each ink in a range of 4-7 μs and the maximum voltage between 40-60 V. A jetting frequency of 1000 Hz was used for all the tests. Before printing, the ink was filtered using a syringe filter with a 700 nm mesh.

Inks were printed on a pre-sintered YSZ/NiO composite produced by tape casting as described elsewhere [51]. The deposition was carried out placing droplets in squared arrangement at different distances in order to achieve an optimum compromise between substrate coverage and minimum overlap. For this specific substrate, the optimal linear droplet density was 500 dots per inch in the x and y directions. In the case of multilayer printing, *ca.* 120 s of drying time was applied between the depositions of consecutive layers. YSZ samples were sintered at three different maximum temperatures in air: 800, 1000 and 1200 °C. The following thermal profile was used: 0 – 600 °C at 0.25 °C min$^{-1}$; dwell for 4 h; 600 – T$_{max}$ °C at 1 °C min$^{-1}$; dwell for 6 h; cool to room temperature at 1.67 °C min$^{-1}$.

The SEM characterization of the sintered samples was carried out with a Zeiss MERLIN scanning electron microscope. Top-view images were taken on carbon coated samples while cross sections were observed after embedding samples in epoxy followed by polishing. The porosity of the sintered samples was estimated by image analysis with the software ImageJ [52].

## 3. Results and discussion

*3.1 Synthesis and inks formulation*

Water dispersions of YSZ nanoparticles were produced in supercritical conditions by heating and pressurizing the water stream at 375 °C and 241 bar in the Promethean reactor, and 397 °C and 270 bar in the DTU reactor. Figure 1 gives an overview of the typical structure of the CHS reactor, and the particles morphology that were produced in the DTU process.

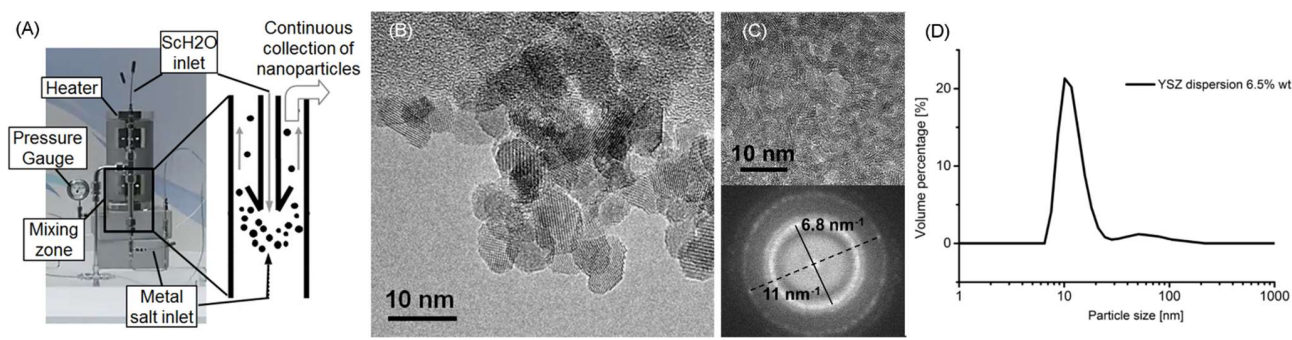

*Figure 1: Sketch of the continuous hydrothermal reactor (Promethean Particles LTD.) with a magnified view of the mixing zone (A), TEM image of the YSZ nanoparticles produced at DTU Energy (B), and Fourier transform of a TEM image at low magnification with randomly distributed particles (C). Figure 1(C) reports the Fourier transform used for calculating the (220) and (111) planes distances. Figure 1(D) shows the DLS analysis of a synthesized YSZ dispersion.*

Particularly, Figure 1(A) shows a sketch of the core components of a continuous hydrothermal reactor operating in supercritical conditions, with a magnified view of the mixing zone. In the configuration showed here, a water stream is first pressurized and heated above its supercritical point and is then continuously mixed with a metal salt stream in a counter-flow fashion [28]. At the interface between the two streams, dehydration and hydrolysis reactions take place leading to the formation of YSZ particles. The morphology for particles synthesized using the DTU reactor was assessed by electron microscopy and is shown in Figure 1(B) and 1(C). The average particle size is 10 nm and each particle consists of a single crystal. Figure 1(C) shows the crystallographic characterization carried out using the fast Fourier transform of an image containing a large number of randomly oriented particles. Two lattice spacings were calculated with values of 2.9(4) and 1.8(1) Å, in good agreement with the crystallographic distances along the [220] and [111] directions of cubic YSZ. TEM characterization especially revealed very similar structures between three different batches that were characterized. Besides the particles in Figure 1, YSZ nanoparticles produced at Promethean by using supercritical water at 375 °C and 241 bar were observed. Despite the use of different CHS conditions, neither morphological nor crystallographic differences were appreciated at the TEM for the material produced in the two reactors.

The ink formulation was carried out bearing in mind the 1-10 range of the printability Z reported by Derby [15] and Reis [37], as already done in our previous works [50]. At the same time, a wider range of Z was also explored for assessing the modifications needed to make a CHS product printable.

YSZ nano-dispersions were synthesized in water and concentrated to a final solid loading of 5.5 ±1 %wt by water evaporation or tangential flow water filtration. Due to the low particle concentration, viscosity ($\eta$) and surface tension ($\sigma$) were comparable with the values for pure water, *i.e.* 1 mPa s and 73 mN m$^{-1}$ respectively. In particular, the as-produced nano-dispersions showed a surface tension of 71 mN m$^{-1}$ and a viscosity of 1.2 mPa s at a shear rate of 1000 s$^{-1}$. These results indicated only a weak impact of the presence of nanoparticles on these properties, leading to a printability value as high as Z = 33. These dispersions showed very little agglomeration and a limited reversible sedimentation after 72h. Figure 1(D) reports the particle size distribution measured by DLS, indicating that the dispersion consisted of 6 – 20 nm particles with a maximum volume percentage at 10 nm, in good agreement with the TEM characterization. Above 30 nm up to 1 μm, only a small and broad peak with a maximum at 50 nm was visible, indicating substantial absence of aggregates that could influence the printing process. The good dispersion of the particles was explained by measuring the ζ-potential. This analysis resulted in a value of 62 ± 2 mV, in the typical range of dispersions with a strong electrostatic stabilization. From this starting composition, inks with different Z values were prepared to investigate the jetting behavior of YSZ nanoparticle-loaded fluids.

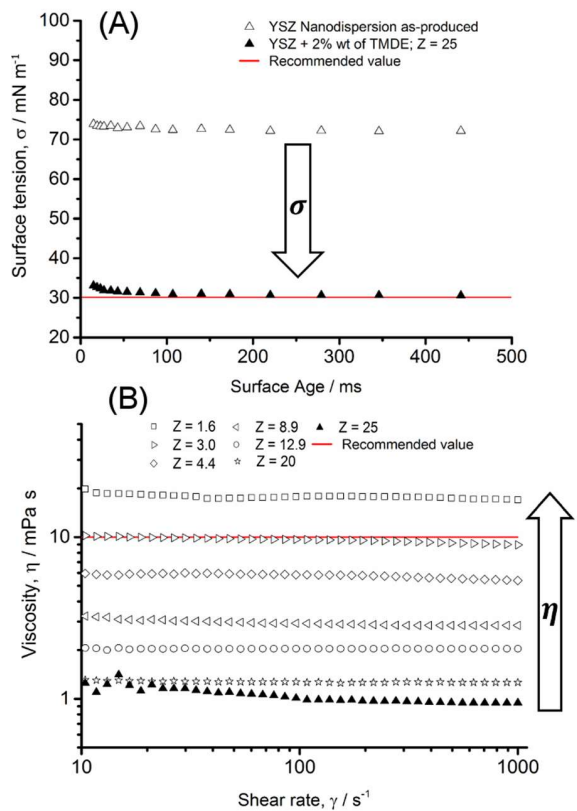

*Figure 2: Rheological optimization of the YSZ nano-dispersions. Measurement of the surface tension before and after the addition of TMDE (A), viscosities at increasing shear rates and resulting Z values at 1000 s$^{-1}$ rate (B). The red line marks the surface tension and viscosity recommended by the printhead manufacturer for an optimal jetting.*

The ink formulation is summarized in Figure 2 and was started with lowering σ to the optimal surface tension recommended by the printhead manufacturer, *i.e.* 30 mN m$^{-1}$. Figure 2(A) indicates that adding 2 %wt of TMDE allowed reaching an optimal value of the surface tension without increasing the viscosity (black triangles in Figure 2(A) and (B)). In the following steps, the surface tension had been kept constant to 30 mN m$^{-1}$ while the viscosity was varied. Since the printability dependency with viscosity and surface tension is $1/\eta$ and $\sqrt{\sigma}$ respectively, we chose to control η. Hence, a larger Z interval can be achieved by smaller modification of the viscosity.

Several YSZ nano-inks with different viscosities were formulated using PVP as an additive after lowering σ with TMDE. Figure 2(B) reports the viscosities of these inks at increasing shearing rates, highlighting a slight self-thinning behavior of the fluids. This weak viscosity decrease is noteworthy because it suggests a limited agglomeration of particles in the fluid, which is unexpected considering the small particle size and their large surface area [16]. The all-wet synthesis of YSZ nanoparticles likely contributed to the prevention of substantial aggregation. Inks having viscosities measured at 1000 s$^{-1}$ ranging from 1.2 to 17 mPa s were prepared in this way (Figure 2(B)). The viscosity plots are labelled with the Z value of each ink, which covers a wide range going from 1.6 to 25. It can be noted that the viscosity recommended by the printhead manufacturer translates into a Z value of about 3.

All these inks were loaded into the cartridges and subsequently mounted on the printer.

*3.2 Jetting behavior and printing*

Jetting tests were conducted to analyze the jetting behavior of the formulated inks. For each ink, the waveform was optimized for reaching the best droplet shape and the most reproducible droplet generation. This always resulted in a trapezoidal electrical pulse where only the duration of the filling ramp and the maximum voltage were found to have a significant effect on the jetting. Therefore, during the waveform optimization, dwell and fall times were kept constant at 10 and 5 μs, respectively. On the other hand, the duration of the filling ramp and the value for the maximum voltage changed for each ink and were comprised between 4-7 μs and 40-60 V, respectively. In particular, for low Z inks shorter filling ramps and higher voltages were required than for high Z inks.

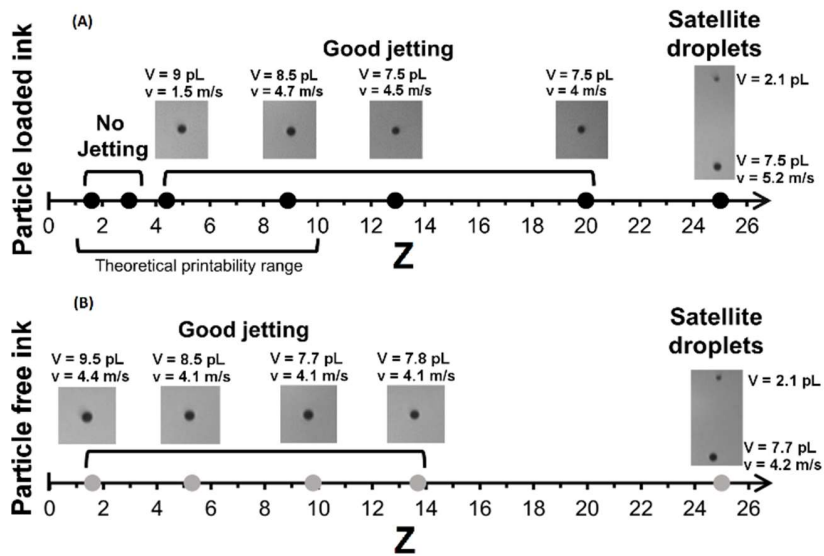

*Figure 3: Jetting behavior of the formulated particle-loaded inks with different Z values (A). Jetting behavior of particle-free inks with similar Z values of the particle-loaded inks (B). Both (A) and (B) report a droplet picture, the droplet volume V and speed v obtained at each Z.*

Figure 3 summarizes the results obtained with several inks, which are displayed here as points defined by different Z values. For particle-loaded inks, Figure 3(A) describes their jetting behavior showing droplet volume, speed and shape obtained with an optimized waveform. To investigate the effect of nanoparticles on the jetting behavior, a similar set of experiments was carried out formulating particle-free inks (Figure 3(B)). Mixtures of water, PVP and TMDE were prepared to explore a printability range similar to the particle-loaded inks, using the same components of their liquid phase.

Interestingly, particle-loaded inks with Z comprised between 4.4 and 20 allowed obtaining single, round shaped droplets. On the opposite at Z = 25 the formation of a satellite droplet was unavoidable, while it was impossible to achieve stable jetting with inks having Z values equal to 1.6 and 3.0 (Figure 3(A)). For particle-free inks, Figure 3(B) shows that inks with low Z, i.e. high viscosity, were jettable and produced good quality droplets. An optimal jetting was observed for Z values from 1.6 to 14, while at Z = 25 satellite droplets were formed. For both inks, the droplet volume followed a similar increasing trend with the fluid viscosity. This last observation is in agreement with Mogalicherla *et al.* [42], but there is no accordance in literature on the relation between Z and drop dimension [53]. In summary, for particle-loaded inks a good printability range was defined by Z values between 4.4 and 20. Fluids with higher viscosities and Z < 4 resulted in no jetting, while for Z > 20 multiple droplet formation was observed. The comparison with particle-free inks indicated that the no-jetting zone found at Z < 4 was not an artifact, but a reproducible behavior induced by the presence of nanoparticles in the fluid.

The good printability range for particle-loaded inks is in disagreement with previous experimental observations and theoretical calculations. In particular, modeling and simulation studies indicated that the Z range for good jetting is 1 < Z < 10 [37]. Deposition of zirconia by inkjet printing has been reported also using reactive inks based on sol-gel chemistry [54] and it had been demonstrated that inks with Z between 1 and 2 were printed successfully. Our experiments with particle-free inks confirmed these results but the addition of nanoparticles avoided jetting for Z < 4. This lowest bound for Z is in accordance with the work of Jang *et al.* who observed a good printability range of 4 < Z < 14 [36]. On the other hand, the top limit of the printability range is higher than the values observed in both theoretical and experimental early studies of inkjet printing [36,37,39] . It is noteworthy that this significant modification of the printability range was caused by a very small nanoparticle concentration. We can thus conclude that the particle nanosize plays a crucial role on the fluid dynamic of the inkjet process. Due to their small size and large surface area, a small quantity of particles with a diameter of 10 nm forms a high amount of solid-liquid interfaces. These interfaces might represent discontinuities into the liquid medium that can affect the transmission of mechanical stresses through the fluid. The addition of larger particles was previously reported to extend the printability range towards higher values. However, a solid loading of 24 %vol leads to an increase of the printability to Z = 16.7 [39], while our results were obtained with a concentration of 0.9 %vol. Our results are in better agreement with the work of Mogalicherla *et al.*, where good jetting was observed at Z as high as 30 using an ink prepared with 1.25 %vol of $Al_2O_3$ nanoparticles [42].

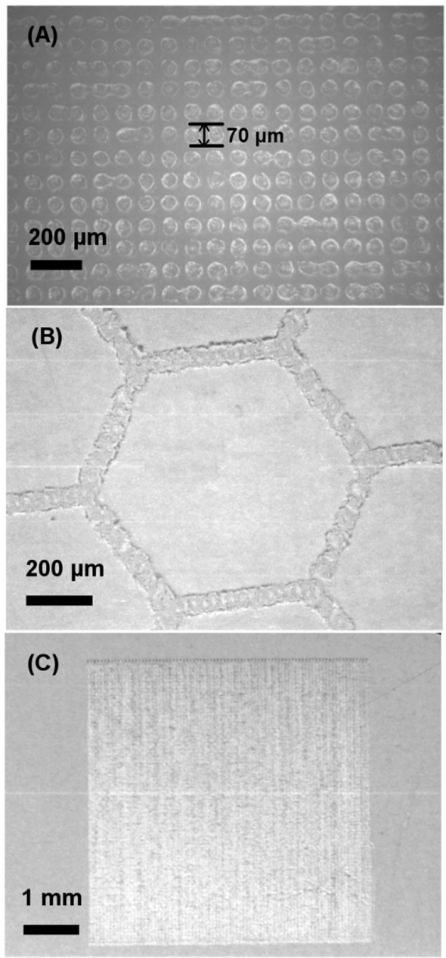

*Figure 4: separated splats (A), honeycomb pattern (B) and continuous squared layer (C) all obtained using the ink with Z = 20 and by overprinting 10 times with a linear density of 500 dots per inch.*

After studying the jettability of the nano-inks, those results were exploited for patterning the substrate with YSZ nanoparticles. Firstly, printing experiments were carried out using the ink formulation with Z = 20, in order to understand the optimal conditions for the fabrication of patterns and continuous layers. In particular, several different dots dispositions were tested to achieve the best overlap between the splats. Indeed, when a droplet impacts on the substrate, it generates a splat with a diameter depending on several parameters: droplet volume and speed, ink viscosity, ink surface tension and surface energy of the substrate. The splat diameter corresponds to the smallest feature that can be printed and therefore determines the highest possible lateral resolution for a specific set of conditions. This value was measured by analyzing the splat diameter of separated dots and it resulted to be 70 μm on a sintered YSZ/NiO composite, see Figure 4(A). The droplet volume and splat diameter indicated that in our inkjet process, it was possible to position less than 0.5 ng of nanoparticles in an area of 0.04 mm$^2$. Avoiding accumulation of material and reaching a smooth surface is important for reducing defects during sintering. The optimal arrangement consisted of dots in a squared disposition, as in Figure 4(A), printed with a linear density of 500 dots per inch in both x and y directions. It was possible to draw complex patterns in these conditions as the honeycomb reported in Figure 4(B) and uniform continuous layers as in Figure 4(C). Due to the dependence of the lateral resolution from the surface energy of the substrate, these optimized parameters needed to be adjusted for each different substrate.

Using these optimized conditions, square shaped samples (4x4 mm$^2$) were deposited for studying the sintering of YSZ nanoparticles in constrained conditions. To this scope, 10 layers were overprinted onto a pre-sintered substrate.

*3.3 Sintering*

A first set of experiments was carried out to investigate the sintering of YSZ nanoparticles in free conditions at 800, 1000 and 1200 °C in air.

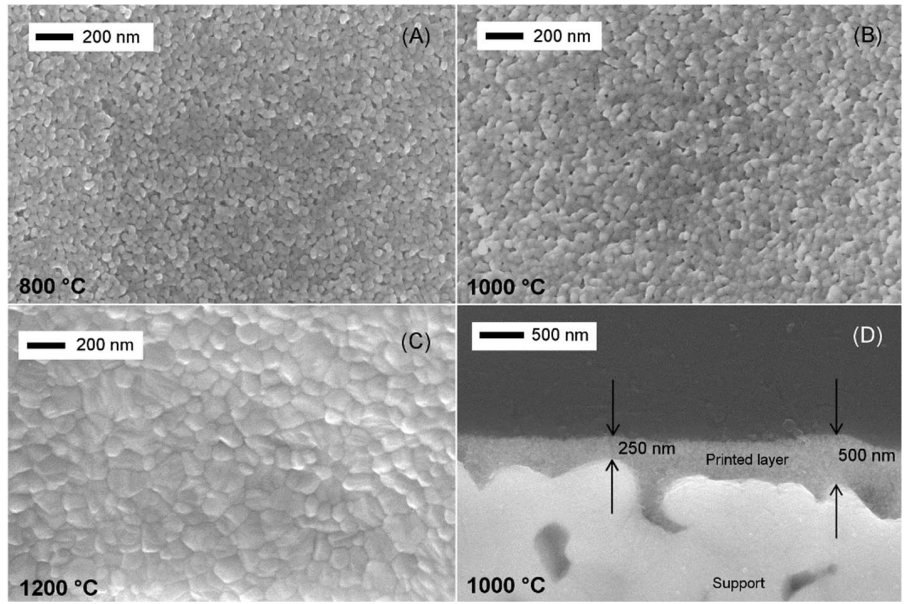

*Figure 5: microstructure of YSZ nanoparticles after free sintering at 800 °C (A), 1000 °C (B), 1200 °C (C) , and cross-section of the printed film after sintering in constrained conditions at 1000 °C (D).*

Figure 5 shows the microstructures obtained in these experiments. Figures 5 A-C show the top view of the free sintering samples at 800 °C, 1000 °C and 1200 °C. These images highlight a tendency of the nanoparticles towards necking and grain growth, which is already visible at 800 °C. At this temperature, sintering was already at an advanced stage as substantial necking of the particles was observed leaving a residual porosity of *ca.* 9 %vol. Grain growth was also measured, as the gain size increased from the starting 10 nm to about 25-30 nm. At 1000 °C, residual porosity was further decreased to *ca.* 2 %vol and particle started coalescing as a consequence of the grain growth. Full densification was reached at 1200 °C, where we observed the formation of crystallites up to 200 nm with flat grain boundaries. This last feature is a typical evidence of the complete relaxation of the material as a result of an intense mass diffusion.

Square shaped layers, as in Figure 4(C), were used for studying the sintering behavior of the material in constrained conditions. Each sample was prepared by printing 10 layers of YSZ nanoparticles on a pre-sintered substrate, so that these films experienced a shrinkage due to the sintering and an opposite stress due to the thermal expansion of the substrate. Figure 5(D) shows the cross section of a sample sintered at 1000 °C. This picture reveals the microstructure of the printed material, which contained extremely fine pores (10 - 40 nm). Similarly to the sample sintered at 1000 °C in free conditions (Figure 5(B)), particles connected into a continuous network with a nearly full densification (estimated residual porosity 2 %vol). The dimension of the superficial porosity was two orders of magnitude higher than the starting particles size. Nonetheless, infiltration through the substrate was not observed and particles formed a continuous layer which covered the whole support. Moreover, the film obtained after sintering at 1000 °C was self-levelling, *i.e.* for ten printed layers its thickness varied between 250 and 500 nm to smooth the substrate morphology. The major reason for this thickness variation is the pronounced roughness of the substrate, which presented pinning features and large cavities. The smoothing effect is visible in Figure 5(D), which highlights the flat top surface of the inkjet printed material in contrast with the irregular substrate. The SEM analysis showed this behavior for all the printed samples. Nanoparticles arranged in such a way that the material adapted to the irregular surface keeping a constant level over pores and asperities.

4. **Conclusions**

We here report for the first time how to transfer YSZ nano-dispersions produced by continuous hydrothermal synthesis in inks for inkjet printing. The all-wet process for the ink preparation avoided a significant aggregation resulting in continuous networks of particles after sintering. Furthermore, the presence of particles with extremely reduced size affected the jetting behavior of the fluids compared to particle-free inks with similar compositions. In particular, nanoparticles hindered the generation of droplets for inks with $Z < 4$, and at the same time prevented the formation of satellite droplets for Z as high as 20. The transfer of CHS products into the inkjet process was completed by optimizing the printing process with the formulated inks. This allowed writing patterns with a lateral resolution of about 70 μm. Inkjet printed YSZ layers were sintered on a porous pre-sintered substrate with a very rough surface showing a beneficial self-levelling effect. This allowed fabricating continuous layers with a thickness between 250 nm – 500 nm depending on the local structure of the surface. Nevertheless, no infiltration of the nanoparticles took place although the porosity of the substrate was orders of magnitude larger than the starting particle size. YSZ nanoparticles showed also a strong tendency toward sintering. In particular, necking and grain growth were already visible at temperatures as low as 800 °C, while full densification was achieved at 1200 °C. A further evidence of the remarkable reactivity of these nanoparticles was given by their sintering in constrained conditions, where the particles were deposited on pre-sintered substrates.

## Acknowledgements

This project has partially received funding from the Fuel Cells and Hydrogen 2 Joint Undertaking under grant agreement No 700266. This Joint Undertaking receives support from the European Union's Horizon 2020 research and innovation program and Hydrogen Europe and N.ERGHY.

# References

[1] G. Li, Z. Tang, Noble metal nanoparticle@metal oxide core/yolk-shell nanostructures as catalysts: recent progress and perspective, Nanoscale. 6 (2014) 3995–4011. doi:10.1039/c3nr06787d.

[2] A. Rogov, Y. Mugnier, L. Bonacina, Harmonic nanoparticles: noncentrosymmetric metal oxides for nonlinear optics, J. Opt. 17 (2015) 33001. doi:10.1088/2040-8978/17/3/033001.

[3] S.T. Kochuveedu, Y.H. Jang, D.H. Kim, A study on the mechanism for the interaction of light with noble metal-metal oxide semiconductor nanostructures for various photophysical applications, Chem. Soc. Rev. 42 (2013) 8467. doi:10.1039/c3cs60043b.

[4] B. Wetzel, P. Rosso, F. Haupert, K. Friedrich, Epoxy nanocomposites - fracture and toughening mechanisms, Eng. Fract. Mech. 73 (2006) 2375–2398. doi:10.1016/j.engfracmech.2006.05.018.

[5] N. Satoh, T. Nakashima, K. Kamikura, K. Yamamoto, Quantum size effect in TiO2 nanoparticles prepared by finely controlled metal assembly on dendrimer templates, Nat. Nanotechnol. 3 (2008) 106–111. doi:10.1038/nnano.2008.2.

[6] A.T. Bell, G.C. Bond, D.T. Thompson, M. Valden, X. Lai, D.W. Goodman, T. Blasko, J.M.L. Nieto, K. Chen, A.T. Bell, E. Iglesia, T. Koyama, T. Komaya, P.L. Gai, M. Weyland, G. Durscher, N.D. Browning, S.J. Pennycook, F. Besenbacker, P.C. Thune, J. Loss, D. Wonter, P.J. Leustra, J.W. Niemantsverdriet, J. Corker, V. Vidal, A. Theolier, J. Thivolle-Cazat, J.-M. Basset, C. Nozkaki, C.G. Lugmair, A.T. Bell, T.D. Tilley, D. Kolb, K.P. De Jong, J.W. Geus, The impact of nanoscience on heterogeneous catalysis., Science. 299 (2003) 1688–91. doi:10.1126/science.1083671.

[7] V. Polshettiwar, R.S. Varma, Green chemistry by nano-catalysis, Green Chem. 12 (2010) 743. doi:10.1039/b921171c.

[8] D.Z. De Florio, V. Esposito, E. Traversa, R. Muccillo, F.C. Fonseca, Master sintering curve for Gd-doped CeO2 solid electrolytes, J. Therm. Anal. Calorim. 97 (2009) 143–147. doi:10.1007/s10973-009-0238-6.

[9] J.A. Glasscock, V. Esposito, S.P. V Foghmoes, T. Stegk, D. Matuschek, M.W.H. Ley, S.

Ramousse, The effect of forming stresses on the sintering of ultra-fine Ce0.9Gd0.1O2-?? powders, J. Eur. Ceram. Soc. 33 (2013) 1289–1296. doi:10.1016/j.jeurceramsoc.2012.12.015.

[10] V. Esposito, E. Traversa, Design of electroceramics for solid oxides fuel cell applications: Playing with ceria, J. Am. Ceram. Soc. 91 (2008) 1037–1051. doi:10.1111/j.1551-2916.2008.02347.x.

[11] E. Olevsky, T.T. Molla, H.L. Frandsen, R. Bjørk, V. Esposito, D.W. Ni, A. Ilyina, N. Pryds, Sintering of multilayered porous structures: Part I-constitutive models, J. Am. Ceram. Soc. 96 (2013) 2657–2665. doi:10.1111/jace.12375.

[12] D.W. Ni, E. Olevsky, V. Esposito, T.T. Molla, S.P. V Foghmoes, R. Bjørk, H.L. Frandsen, E. Aleksandrova, N. Pryds, Sintering of multilayered porous structures: Part II-experiments and model applications, J. Am. Ceram. Soc. 96 (2013) 2666–2673. doi:10.1111/jace.12374.

[13] D.W. Ni, C.G. Schmidt, F. Teocoli, A. Kaiser, K.B. Andersen, S. Ramousse, V. Esposito, Densification and grain growth during sintering of porous Ce0.9Gd0.1O1.95 tape cast layers: A comprehensive study on heuristic methods, J. Eur. Ceram. Soc. 33 (2013) 2529–2537. doi:10.1016/j.jeurceramsoc.2013.03.025.

[14] T.H.J. Van Osch, J. Perelaer, A.W.M. De Laat, U.S. Schubert, Inkjet printing of narrow conductive tracks on untreated polymeric substrates, Adv. Mater. 20 (2008) 343–345. doi:10.1002/adma.200701876.

[15] B. Derby, Inkjet printing ceramics : From drops to solid, 31 (2011) 2543–2550. doi:10.1016/j.jeurceramsoc.2011.01.016.

[16] S. Kango, S. Kalia, A. Celli, J. Njuguna, Y. Habibi, R. Kumar, Surface modification of inorganic nanoparticles for development of organic-inorganic nanocomposites - A review, Prog. Polym. Sci. 38 (2013) 1232–1261. doi:10.1016/j.progpolymsci.2013.02.003.

[17] A. Kosmala, Q. Zhang, R. Wright, P. Kirby, Development of high concentrated aqueous silver nanofluid and inkjet printing on ceramic substrates, Mater. Chem. Phys. 132 (2012) 788–795. doi:10.1016/j.matchemphys.2011.12.013.

[18] A. Kosmala, R. Wright, Q. Zhang, P. Kirby, Synthesis of silver nano particles and fabrication of aqueous Ag inks for inkjet printing, Mater. Chem. Phys. 129 (2011) 1075–1080. doi:10.1016/j.matchemphys.2011.05.064.

[19] M. Arin, P. Lommens, S.C. Hopkins, G. Pollefeyt, J. Van der Eycken, S. Ricart, X. Granados, B. a Glowacki, I. Van Driessche, Deposition of photocatalytically active TiO$_2$ films by inkjet printing of TiO$_2$ nanoparticle suspensions obtained from microwave-assisted hydrothermal synthesis., Nanotechnology. 23 (2012) 165603. doi:10.1088/0957-4484/23/16/165603.

[20] S. Lee, T. Boeltken, A.K. Mogalicherla, U. Gerhards, P. Pfeifer, R. Dittmeyer, Inkjet printing of porous nanoparticle-based catalyst layers in microchannel reactors, Appl. Catal. A Gen. 467 (2013) 69–75. doi:10.1016/j.apcata.2013.07.002.

[21] R. Cherrington, D.J. Hughes, S. Senthilarasu, V. Goodship, Inkjet-Printed TiO2 Nanoparticles from Aqueous Solutions for Dye-Sensitized Solar Cells (DSSCs), Energy Technol. 3 (2015) 866–870. doi:10.1002/ente.201500096.

[22] P.J. Wojcik, L. Santos, L. Pereira, R. Martins, E. Fortunato, Tailoring nanoscale properties of tungsten oxide for inkjet printed electrochromic devices, Nanoscale. 7 (2015) 1696–1708. doi:10.1039/C4NR05765A.

[23] I. Fasaki, K. Siamos, M. Arin, P. Lommens, I. Van Driessche, S.C. Hopkins, B.A. Glowacki, I. Arabatzis, Ultrasound assisted preparation of stable water-based nanocrystalline TiO2 suspensions for photocatalytic applications of inkjet-printed films, Appl. Catal. A Gen. 411–412 (2012) 60–69. doi:10.1016/j.apcata.2011.10.020.

[24] P.D. Angelo, R. Kronfli, R.R. Farnood, Synthesis and inkjet printing of aqueous ZnS:Mn nanoparticles, J. Lumin. 136 (2013) 100–108. doi:10.1016/j.jlumin.2012.10.043.

[25] P. Zielke, Y. Xu, S.B. Simonsen, P. Norby, R. Kiebach, Simulation, design and proof-of-concept of a two-stage continuous hydrothermal flow synthesis reactor for synthesis of functionalized nano-sized inorganic composite materials, J. Supercrit. Fluids. 117 (2016) 1–12. doi:10.1016/j.supflu.2016.06.008.

[26] T. Adschiri, K. Kanazawa, K. Arai, Rapid and Continuous Hydrothermal Crystallization

of Metal Oxide Particles in Supercritical Water, J. Am. Ceram. Soc. 75 (1992) 1019–1022. doi:10.1111/j.1151-2916.1992.tb04179.x.

[27] T. Adschiri, Y. Hakuta, K. Sue, K. Arai, Hydrothermal synthesis of metal oxide nanoparticles at supercritical conditions, J. Nanoparticle Res. 3 (2001) 227–235. doi:10.1023/A:1017541705569.

[28] E. Lester, P. Blood, J. Denyer, D. Giddings, B. Azzopardi, M. Poliakoff, Reaction engineering: The supercritical water hydrothermal synthesis of nano-particles, J. Supercrit. Fluids. 37 (2006) 209–214. doi:10.1016/j.supflu.2005.08.011.

[29] R.I. Brief, R. Summaries, Sustainable Hydro thermal Manufacturing o f Nano materials Pro ject details, (2016) 1–5.

[30] M. Černá, M. Veselý, P. Dzik, C. Guillard, E. Puzenat, M. Lepičová, Fabrication, characterization and photocatalytic activity of TiO2 layers prepared by inkjet printing of stabilized nanocrystalline suspensions, Appl. Catal. B Environ. 138–139 (2013) 84–94. doi:10.1016/j.apcatb.2013.02.035.

[31] A. Lee, K. Sudau, K.H. Ahn, S.J. Lee, N. Willenbacher, Optimization of experimental parameters to suppress nozzle clogging in inkjet printing, Ind. Eng. Chem. Res. 51 (2012) 13195–13204. doi:10.1021/ie301403g.

[32] X. Wang, W.W. Carr, D.G. Bucknall, J.F. Morris, High-shear-rate capillary viscometer for inkjet inks, Rev. Sci. Instrum. 81 (2010). doi:10.1063/1.3449478.

[33] J.E. Fromm, Numerical Calculation of the Fluid Dynamics of Drop-on-Demand Jets, IBM J. Res. Dev. 28 (1984) 322–333. doi:10.1147/rd.283.0322.

[34] Q. Xu, O.A. Basaran, Computational analysis of drop-on-demand drop formation, Phys. Fluids. 19 (2007). doi:10.1063/1.2800784.

[35] J. Eggers, E. Villermaux, Physics of liquid jets, Reports Prog. Phys. 71 (2008) 36601. doi:10.1088/0034-4885/71/3/036601.

[36] D. Jang, D. Kim, J. Moon, Influence of fluid physical properties on ink-jet printability, Langmuir. 25 (2009) 2629–2635. doi:10.1021/la900059m.

[37] N. Reis, B. Derby, Ink Jet Deposition of Ceramic Suspensions: Experiments of Droplet Formation, Mater. Res. 625 (2000) 117–122. doi:dx.doi.org/10.1557/PROC-624-65.

[38] C. Gadea, D. Marani, V. Esposito, Nucleophilic stabilization of water-based reactive ink for titania-based thin film inkjet printing, J. Phys. Chem. Solids. 101 (2017) 10–17. doi:10.1016/j.jpcs.2016.10.004.

[39] E. Özkol, J. Ebert, R. Telle, An experimental analysis of the influence of the ink properties on the drop formation for direct thermal inkjet printing of high solid content aqueous 3Y-TZP suspensions, J. Eur. Ceram. Soc. 30 (2010) 1669–1678. doi:10.1016/j.jeurceramsoc.2010.01.004.

[40] M. Bienia, M. Lejeune, M. Chambon, V. Baco-Carles, C. Dossou-Yovo, R. Noguera, F. Rossignol, Inkjet printing of ceramic colloidal suspensions: Filament growth and breakup, Chem. Eng. Sci. 149 (2016) 1–13. doi:10.1016/j.ces.2016.04.015.

[41] M. Corcione, Empirical correlating equations for predicting the effective thermal conductivity and dynamic viscosity of nanofluids, Energy Convers. Manag. 52 (2011) 789–793. doi:10.1016/j.enconman.2010.06.072.

[42] A.K. Mogalicherla, S. Lee, P. Pfeifer, R. Dittmeyer, Drop-on-demand inkjet printing of alumina nanoparticles in rectangular microchannels, Microfluid. Nanofluidics. 16 (2014) 655–666. doi:10.1007/s10404-013-1260-3.

[43] H. Huang, M. Nakamura, P. Su, R. Fasching, Y. Saito, F.B. Prinz, High-Performance Ultrathin Solid Oxide Fuel Cells for Low-Temperature Operation, J. Electrochem. Soc. 154 (2007) B20. doi:10.1149/1.2372592.

[44] J.M. Vohs, R.J. Gorte, High-performance SOFC cathodes prepared by infiltration, Adv. Mater. 21 (2009) 943–956. doi:10.1002/adma.200802428.

[45] F. Matei, C. Jiménez-Borja, J. Canales-Vázquez, S. Brosda, F. Dorado, J.L. Valverde, D. Ciuparu, Enhanced electropromotion of methane combustion on palladium catalysts deposited on highly porous supports, Appl. Catal. B Environ. 132–133 (2013) 80–89. doi:10.1016/j.apcatb.2012.11.011.

[46] N.M. Farandos, L. Kleiminger, T. Li, A. Hankin, G.H. Kelsall, Three-dimensional Inkjet

Printed Solid Oxide Electrochemical Reactors. I. Yttria-stabilized Zirconia Electrolyte, Electrochim. Acta. 213 (2016) 324–331. doi:10.1016/j.electacta.2016.07.103.

[47] J.C. Ruiz-Morales, A. Tarancón, J. Canales-Vázquez, J. Méndez-Ramos, L. Hernández-Afonso, P. Acosta-Mora, J.R. Marín Rueda, R. Fernández-González, Three dimensional printing of components and functional devices for energy and environmental applications, Energy Environ. Sci. (2017) 846–859. doi:10.1039/C6EE03526D.

[48] Q. Zhu, B. Fan, Low temperature sintering of 8YSZ electrolyte film for intermediate temperature solid oxide fuel cells, Solid State Ionics. 176 (2005) 889–894. doi:10.1016/j.ssi.2004.12.010.

[49] Pall, Introduction to Tangential Flow Filtration for Laboratory and Process Development Applications, (2014) 1–13.

[50] V. Esposito, C. Gadea, J. Hjelm, D. Marani, Q. Hu, K. Agersted, S. Ramousse, S.H. Jensen, Fabrication of thin yttria-stabilized-zirconia dense electrolyte layers by inkjet printing for high performing solid oxide fuel cells, J. Power Sources. 273 (2015) 89–95. doi:10.1016/j.jpowsour.2014.09.085.

[51] A. Hauch, M. Mogensen, A. Hagen, Ni/YSZ electrode degradation studied by impedance spectroscopy - Effect of p(H2O), Solid State Ionics. 192 (2011) 547–551. doi:10.1016/j.ssi.2010.01.004.

[52] C.A. Schneider, W.S. Rasband, K.W. Eliceiri, NIH Image to ImageJ: 25 years of image analysis, Nat. Methods. 9 (2012) 671–675. doi:10.1038/nmeth.2089.

[53] D. Kuscer, G. Stavber, G. Trefalt, M. Kosec, Formulation of an aqueous titania suspension and its patterning with ink-jet printing technology, J. Am. Ceram. Soc. 95 (2012) 487–493. doi:10.1111/j.1551-2916.2011.04876.x.

[54] C. Gadea, Q. Hanniet, A. Lesch, D. Marani, S.H. Jensen, V. Esposito, Aqueous metal-organic solutions for YSZ thin film inkjet deposition, J. Mater. Chem. C. (2017). doi:10.1039/C7TC01879G.